# Controlling the toroidal excitations in metamaterials for high-Q response

Yuancheng Fan,[1,*] Fuli Zhang,[1] Quanhong Fu,[1] Zeyong Wei,[2,3] and Hongqiang Li[2,3]

[1]*Key Laboratory of Space Applied Physics and Chemistry, Ministry of Education and Department of Applied Physics, School of Science, Northwestern Polytechnical University, Xi'an 710129, China*
[2]*Key Laboratory of Advanced Micro-structure Materials (MOE) and Department of Physics, Tongji University, Shanghai 200092, China*
[3]*The Institute of Dongguan-Tongji University, Dongguan 523808, Guangdong, China*
*phyfan@nwpu.edu.cn*

**Abstract:** The excitation of toroidal multipoles in metamaterials was investigated for high-*Q* response at a subwavelength scale. In this study, we explored the optimization of toroidal excitations in a planar metamaterial comprised of asymmetric split ring resonators (ASRRs). It was found that the scattering power of toroidal dipole can be remarkably strengthened by adjusting the characteristic parameter of ASRRs: asymmetric factor. Interestingly, the improvement in toroidal excitation accompanies increment on the *Q*-factor of the toroidal metamaterial; it is shown that both the scattering power of toroidal dipole and the *Q*-factor were increased more than one order by changing the asymmetric factor of ASRRs. The optimization in excitation of toroidal multipole provide opportunity to further increase the *Q*-factor of metamaterial and boost light-matter interactions at the subwavelength scale for potential applications in low-power nonlinear processing, and sensitive photonic applications.

## 1. Introduction

Optically resonant modes are of fundamental importance in realizing enhanced light-matter interactions for their coupling ability with light fields in both the spatial and time domain. For the spatial domain coupling, it is highly desirable to localize light in a possibly small volume for strong fields. Conventional dielectric microcavities have been reduced from the "classical" regime to "quantum" regime where wave effects become dominant [1]. And plasmonic excitations in metallic structures have been exploited for realizing light localization at deep subwavelength scale to join together the well-developed nanoelectronic technology [2]. While for the time domain coupling, a dimensionless physical quantity, quality factor ($Q$-factor), is always adopted for measuring energy damping relative to the stored energy of the oscillating, i.e., the photon lifetime of a resonant mode.

Metamaterials and metasurfaces [3-8], which allow light interacts with medium in the resonant mode during its slow damping, are promising for the enhancement of light-matter interactions at a subwavelength scale. Various plasmonic metamaterials have been proposed for achieving high $Q$-factor response, for example, the trapped-mode, which is weakly coupled to free space, was suggested to be excited by introducing symmetry breaking in the shape of structural elements for realizing sharp spectral response [9]. The trapped mode was later demonstrated in similar metamaterials and the resulted high-$Q$ resonance [10, 11] was utilized for applications in ultrasensitive spectroscopy and sensing [12], high-$Q$ filtering [13, 14], optical switching [15], nonlinear optics [16], and lasing spacers [17]. Recently, the excitation of toroidal moment [18] in felicitously designed metamaterial was suggested as a new solution for achieving high-$Q$ resonance [19, 20]. The toroidal excitations are different from electrical and magnetic excitations in traditional multipole expansions [21, 22], it was found that higher-$Q$ response (in comparison with the original magnetic dipolar mode) in a metamaterial can be achieved by taking advantages from the weak free-space coupling of the toroidal dipolar excitation [23-25]. Later, the toroidal excitations were studied in oligomer nanocavities [26], coupled nanodisks [27-29], coupled dielectric/metallic rods [30, 31]. The exotic toroidal response was also investigated for optical manipulations in polarization control [32-35], high-$Q$ subwavelength cavities [36-40], anapole excitation/invisibility [41-44], and optical force [45].

In this paper, we propose to achieve higher $Q$-factor in a planar metamaterial by controlling the toroidal excitations. The unit cell, also called as meta-molecule, of the metamaterial is comprised of four asymmetric split ring resonators (ASRRs). The magnetic resonance of the ASRRs can be excited by electromagnetic wave with polarization direction perpendicular to the gap of the ASRRs. The near-field coupling [23, 46-49] of these magnetic resonators will induce a toroidal moment dominated fundamental resonance in the planar metamaterial. In the study, it is found that the excitation of toroidal mode in a metamaterial can be significantly enhanced, and in the meanwhile, the $Q$-factor of the planar metamaterial can be further improved through the control of toroidal excitations. Furthermore, the fundamental resonance is of Fano-lineshape and the modulation depth of the resonance is nearly unity, which makes the toroidal metamaterials suitable for possible applications in ultra-sensitive sensors. Toroidal geometry together with the Fano resonance made high-$Q$ response will have enormous potential applications in low-threshold lasing, cavity quantum electrodynamics and nonlinear processing.

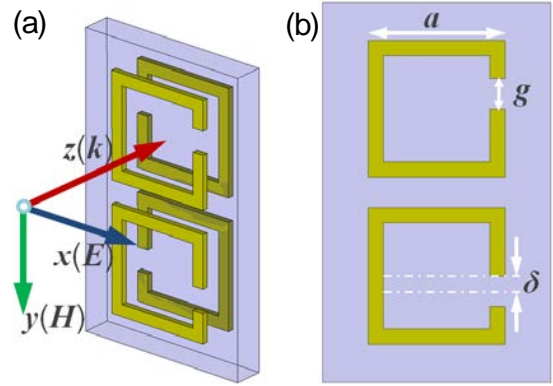

Fig. 1. (a) Schematic of a meta-molecule of the toroidal metamaterial, the metamaterial is illustrated by normal incident wave with the electric polarization is along $x$-axis. (b) Top view of the meta-molecule with the geometric size being labeled by white arrows and black letters.

## 2. Results and discussions

The meta-molecule of the designed toroidal metamaterial is presented in Fig. 1, as can be seen from the figure, the metamaterial is comprised of two patterned metallic layers and a dielectric spacer layer, the metallic layers are patterned into arrays of asymmetric split ring resonators (ASRRs). Four ASRRs are properly arranged for controlling the structural symmetry of the meta-molecule: pairs of ASRRs on same layers are of mirror symmetry about the $xz$-plane; pairs of ASRRs on different layers are of 2-fold rotational symmetry about the $y$-axis. The lattice constants along $x$- and $y$- directions are 5 mm and 10 mm, the permittivity of the 0.8 mm-thick dielectric spacer is 2.55. The width of metal strip is 0.5 mm, the outer width of the ASRRs is $a = 4$ mm, the distance between the ASRRs is 0.8 mm. The asymmetry of the ASRRs was introduced in the square rings by opening gaps of size $g = 1$ mm positioned $\delta$ from the center of the rings, the undetermined asymmetric factor $\delta$ is crucial in the control of the toroidal excitations. We performed all numerical calculations with a finite-difference-in-time-domain (FDTD) [50] electromagnetic solver [51]. And we note that throughout our study the metamaterials are illuminated by $x$-polarized (the electric field $E$ is along $x$-axis) electromagnetic waves as illustrated in Fig. 1(a).

The ASRRs in the toroidal molecule are in close proximity, it is well known that the mutual interactions or couplings [52] between these building blocks play determinative role in their collaborative response to incident fields. The couplings between the ASRRs in both the in-plane pair and the stacked pair were studied in our previous work [36]. The in-plane ASRR pairs were designed of mirror symmetry with respect to the $xz$-plane. The resonant frequency of the paired ASRRs is the same as that of ASRRs. Different from the metamaterials with period ASRRs, the mirror symmetry ensures the paired ASRRs strongly coupled to the incident waves, which makes the trapped magnetic mode of ASRRs can be well excited. To mimic the head-to-tail shaped magnetic morphology of toroidal dipolar excitation in the bulk metamaterial, two layers of metallic ASRRs were stacked to couple the magnetic mode in ASRRs, the paired ASRRs on top and lower layers in the meta-molecule was designed to be of 2-fold rotational symmetry with respect to the *y*-axis. This vertical coupling configuration provides a remarkable overlap between the magnetic dipoles of ASRRs on top and lower layers, indicating strong magnetic coupling. The coupled magnetic modes in ASRRs show in-phase and out-of phase states at low and high frequencies, respectively. The high-frequency resonance is determined by the out-of-phase vertical coupling (not shown here). Since we mainly focused on the toroidal excitation in this paper, we show in Fig. 2 the transmission around the low-frequency resonance and the on-resonance local field distribution, the transmission near the resonant frequency is zoomed in the inset of Fig. 2a. The low-frequency resonance is determined by the in-phase vertical coupling [see color map of the *z*-component magnetic field in Fig. 2(b)] together with horizontal coupling under normal electromagnetic incidence made a circumfluent magnetic field or magnetic vortex, which is exactly the picture of toroidal dipolar [22, 23] which shows a subwavelength electromagnetic localization within dielectric substrate in a toroidal configuration.

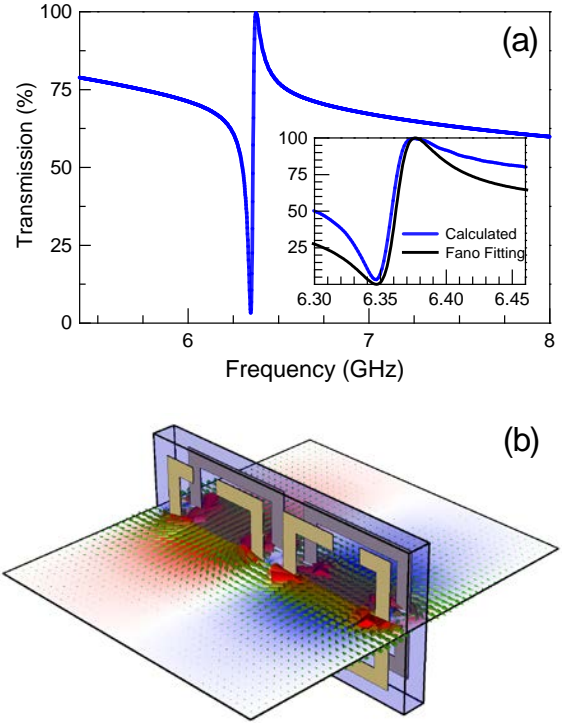

Fig. 2. (a) Calculated spectrum of a designed planar metamaterial with toroidal moment dominated Fano resonance. The asymmetric parameter was set to be: -0.4 mm. The inset shows the zoomed transmission near the resonant frequency. (b) Calculated spatial distributions of vectorial magnetic field over the color map of *z*-component magnetic field on the central plane of a meta-molecule at the peak frequency of the transmission spectrum.

To quantitatively evaluate the contributions on electromagnetic scattering of the multipoles, we adopted the current density formalization [23] by integrating spatial distributed fields in a meta-molecule. In the current density formalization, the radiating power of induced multipoles can be calculated with the formula

$$I = \frac{2\omega^4}{3c^3}|\boldsymbol{P}|^2 + \frac{2\omega^4}{3c^3}|\boldsymbol{M}|^2 + \frac{2\omega^6}{3c^5}|\boldsymbol{T}|^2 + \frac{\omega^6}{20c^5}M_{\alpha\beta}M_{\alpha\beta} + \cdots, \quad (1)$$

with the electric dipole moment being

$$\boldsymbol{P} = \frac{1}{i\omega}\int \boldsymbol{j}\, d^3r, \quad (2)$$

the magnetic dipole moment being

$$\boldsymbol{M} = \frac{1}{2c}\int (\boldsymbol{r}\times\boldsymbol{j})\, d^3r, \quad (3)$$

the toroidal dipole moment being

$$\boldsymbol{T} = \frac{1}{10c}\int \left[(\boldsymbol{r}\cdot\boldsymbol{j})\boldsymbol{r} - 2r^2\boldsymbol{j}\right] d^3r, \quad (4)$$

and the magnetic quadrupole moment being

$$M_{\alpha\beta} = \frac{1}{3c}\int \left[(\boldsymbol{r}\times\boldsymbol{j})_\alpha r_\beta + (\boldsymbol{r}\times\boldsymbol{j})_\beta r_\alpha\right] d^3r, \quad (5)$$

in which $\boldsymbol{j}$ is the current density.

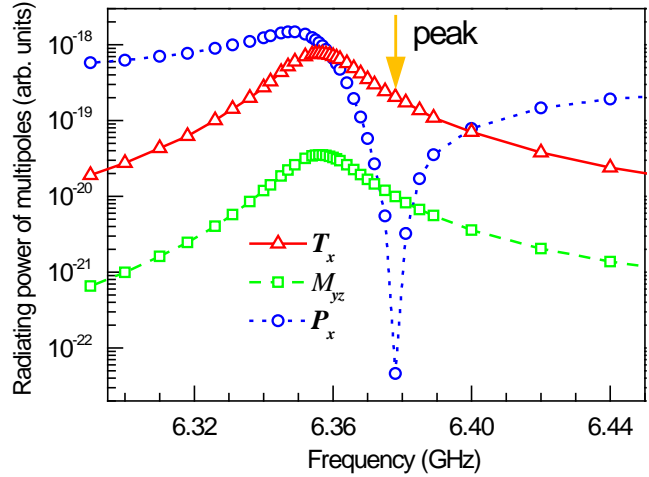

Fig. 3. The radiating power for induced multipoles of the planar toroidal metamaterial: toroidal dipole (red solid curve), magnetic quadrupole (green dashed curve), and electric dipole (blue dotted curve).

The radiating powers of the induced electric dipoles, magnetic dipoles, toroidal dipoles, electric and magnetic quadrupoles were calculated in simulation. We first study the radiating power of the multipoles for the metamaterial with asymmetric parameter $\delta = -0.4$ mm [dashed black lines, the minus sign (-) means the gap is away from the center of the meta-molecule]. For simplicity, we present in Fig. 3 the three mainly contributed multipoles of the metamaterials, including: electric dipole ($\boldsymbol{P}_x$), toroidal dipole ($\boldsymbol{T}_x$), and magnetic quadrupole

($M_{yz}$). We note that the radiating powers of all other components of the multipoles are several orders lower in comparing with the three mainly contributed multipoles.

The electric dipole $P_x$ shows strong scattering in the whole frequency band, which reveal that the metamaterial is excited by electric component of the incoming wave. The toroidal dipole $T_x$ shows considerable scattering around the resonance, the *x*-component of the toroidal dipole moment plays dominative role at the transmission peak frequency 6.38 GHz (marked with arrow in Fig.3) in comparison with other multipoles. The scattering of magnetic magnetic quadrupole $M_{yz}$ shows similar frequency dependency to the toroidal dipole $T_x$, however, the radiating intensity of toroidal dipole $T_x$ is twenty times bigger than that of the magnetic quadrupole $M_{yz}$. It is noteworthy that magnetic quadrupole $M_{yz}$ is the only component with comparable radiating contribution as toroidal dipole, which should be due to their similar coupling configuration of the magnetic dipoles on the *yz*-plane.

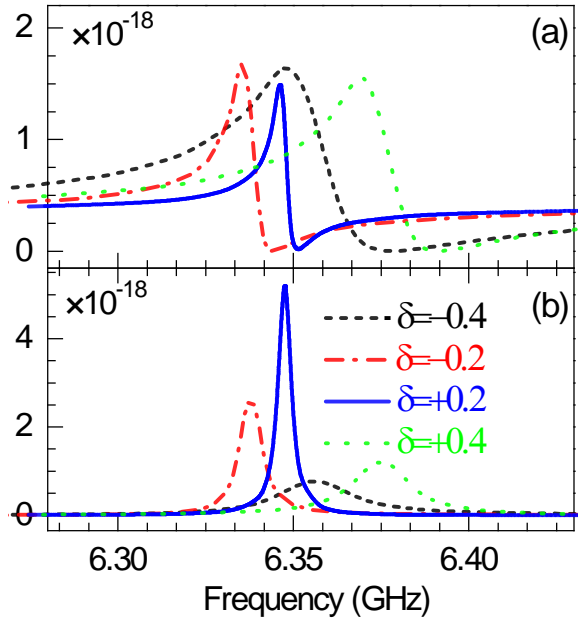

Fig. 4. The radiating power for induced multipoles of the planar toroidal metamaterial with different asymmetric parameter $\delta$: electric dipole (a), toroidal dipole (b).

The meta-molecule of the toroidal metamaterial is constructed with ASRRs, and the unusual property of the ASRR is the high-*Q* trapped mode originating from structural symmetry broken. The asymmetric parameter is crucial in the trap mode of ASRRs which is the fundamental element in the formation of toroidal mode. We thus investigated the influence of asymmetric parameters on the toroidal excitation in metamaterial. Figure 4 shows that the scattering power for induced electric and magnetic dipole is only slightly changed when the asymmetric parameter changes. But differently, the scattering power of toroidal dipole changes significantly along with the change of asymmetric parameter: the toroidal moment is stronger with small asymmetric parameters and when the gap is close to the center of toroidal meta-molecule. The trap mode of the ASRR is sensitive to the asymmetric parameter, an ASRR with small asymmetric factor is weakly coupled to external stimuli and shows stronger binding in time domain, then the excited in-phase magnetic mode, or toroidal mode becomes even more strong compared with bigger asymmetric factor.

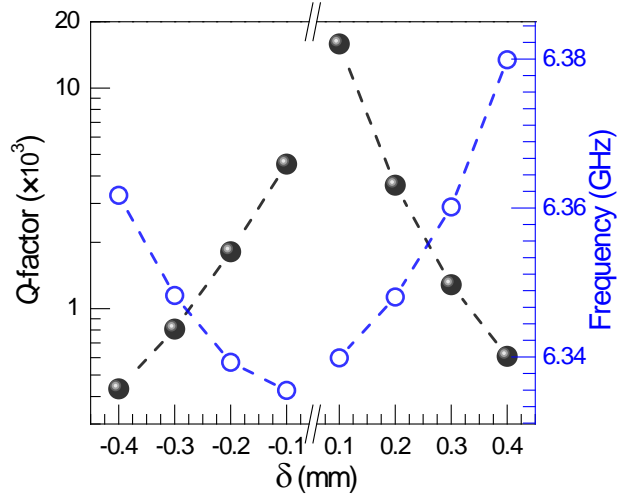

Fig. 5. The resonant frequencies (blue circles) and *Q*-factors (black spheres) of the toroidal metamaterials with asymmetric parameter $\delta$ changing from -0.4 mm to +0.4 mm.

Intuitively, the improvement in exciting the toroidal dipole will be beneficial to achieve higher-*Q* response in the metamaterials for that both the higher-*Q* with smaller asymmetry and the enhanced toroidal excitations are promising in binding electromagnetic waves. The *Q*-factor of a symmetric Lorentzian resonance can be calculated from the ratio of central frequency and FWHM, which is quite convenient. In the meanwhile, the *Q*-factors of asymmetric Fano resonance [53, 54] were always got by fitting the spectrum with a Fano formula

$$I \propto \frac{(F\gamma + \omega - \omega_0)^2}{(\omega - \omega_0)^2 + \gamma^2}, \quad (6)$$

where $\omega_0$ and $\gamma$ represent the position and width of the Fano resonance, and $F$ is the Fano parameter [54]. There are two ways people define *Q*-factors: i) Energy stored in resonator/energy lost per cycle; ii) Resonant frequency/full-width-at-half-maximum. Here, we propose to extract the *Q*-factor from characteristic frequencies (dip and peak frequencies: $f_d$ and $f_p$) of the Fano resonance, which can be understood as the counterpart to the latter in Fano-resonance. The Fano-shaped asymmetric resonance originates from interference between a sharp resonance (discrete state) and the continuum (background). The resonant region lies between the dip and peak frequencies, and $f_d$ and $f_p$ are corresponding to the destructive and constructive interferences, respectively. The central frequency of Fano resonance would be $(f_d + f_p)/2$, and the FWHM would be $|f_d - f_p|/2$, then the *Q*-factor of a Fano resonance can be got in compare with the Lorentzian resonance

$$Q = \frac{f_d + f_p}{|f_d - f_p|} \quad (7)$$

Figure 5 plots the resonant frequencies (blue circles) and calculated $Q$-factors (black spheres) of the toroidal metamaterials with different asymmetric parameter $\delta$ (changing from -0.4 mm to +0.4 mm). The calculated $Q$-factors using Eq. 7 agree well with fitted results using the Fano formula (Eq. 6). Taking the case in Fig. 2 (with $\delta = -0.4$ mm) as an example, we plot in inset of Fig. 2a the fitted spectrum using Eq. 6, the parameters were set as: $F = 1$, $\omega_0/2\pi \approx 6.36 \text{ GHz}$, and $\gamma/2\pi \approx 14.7 \text{ MHz}$. The $Q$-factor $\omega_0/\gamma \approx 432$ is the same as the result calculated from Eq. 7. We can clearly see that $Q$-factor of the toroidal metamaterial increases significantly with small asymmetric parameters and when the gap is close to the center of toroidal meta-molecule, the Q-factor of a metamaterial was increased to 15849 ($\delta = 0.1$ mm) from 432 ($\delta = -0.4$ mm). Compare the scattering of toroidal dipole [in Fig. 4(b)] and $Q$-factor of toroidal mode, we can find that the variation tendency of $Q$-factor is identical to the scattering power of toroidal excitations, which confirms the prediction that the $Q$-factor of toroidal metamaterial can be raised by optimize the excitation of toroidal dipole. Additional numerical studies confirmed that the improvement in $Q$-factor through the optimization of the toroidal excitation is also valid in case of toroidal metamaterial with realistic metal and commercial microwave substrate.

## 3. Conclusion

In summary, we studied the toroidal excitation in a planar metamaterial comprised of ASRRs, it is shown that the toroidal excitation in a planar metamaterial can be controlled by tuning the asymmetric factor of ASRRs, and the enhancement in toroidal excitation can remarkably improve the $Q$-factor of the toroidal metamaterial. The high-$Q$ feature of the toroidal metamaterial and the further improvement on high-$Q$ through geometric optimization may be beneficial to enhance light-matter interactions at a subwavelength scale for low-power nonlinear processing and sensitive photonic application.

## Funding